\documentclass[
reprint,
superscriptaddress,
 showpacs,
 amsmath,amssymb,
 aps,
pre, 
floatfix,
]{revtex4-2}

\usepackage{amsmath,amssymb}
\usepackage{physics}
\usepackage{graphicx, color}
\usepackage{dcolumn}
\usepackage{bm}
\usepackage{bbm}
\usepackage{hyperref}
\usepackage[dvipsnames]{xcolor}
\hypersetup{
setpagesize=false,
 bookmarksnumbered=true,%
 bookmarksopen=true,%
 colorlinks=true,%
 linkcolor=Blue,
 citecolor=Blue,
 urlcolor = Blue,
}
\usepackage{pgffor} 

\makeatletter
\AtBeginDocument{\let\LS@rot\@undefined}
\makeatother

\newif\ifarXiv
\arXivtrue 

\newcommand{\RM}[1]{\mathrm{#1}}

\begin{document}

\preprint{APS/123-QED}

\title{Inferring the dynamics of glass-forming liquids from static structure across thermal states}

\author{Hidemasa Bessho}
\email{bessho.hidemasa@gmail.com}
\affiliation{Department of Physics, Nagoya University, Nagoya 464-8602, Japan}

\author{Takeshi Kawasaki}
\affiliation{D3 Center, The University of Osaka, Toyonaka, Osaka 560-0043, Japan}
\affiliation{Department of Physics, The University of Osaka, Toyonaka, Osaka 560-0043, Japan}

\author{Hayato Shiba}
\email{shiba@gsis.u-hyogo.ac.jp}
\affiliation{Graduate School of Information Science, University of Hyogo, Kobe 650-0047, Japan}

\date{\today}

\begin{abstract}
In this study, we demonstrate the generalizability of graph neural networks in predicting the dynamic heterogeneity of model glass-forming liquids across different temperatures. 
While previous approaches have often been limited to making predictions at the specific temperatures used during training, we find that our proposed framework -- T-BOTAN --  enables interpolation to temperatures not included in the training set.
We show that the dynamical behavior, the associated four-point correlations, and even the macroscopic temperature can be estimated with sufficient accuracy solely from static particle configurations at untrained temperatures. 
These results suggest that static configurations encode not only local structural features driving  dynamic heterogeneity but also fundamental thermodynamic information.
\end{abstract}

\maketitle

\section{Introduction}
When a liquid is cooled sufficiently rapidly to avoid crystallization, it enters a supercooled state and eventually becomes an amorphous solid.
This transformation, known as the glass transition, is characterized by a dramatic slowdown of structural relaxation, yet its microscopic origin remains debated~\cite{angell1995science,Debenedetti2001nat}.
With this slowdown, particle motion becomes increasingly heterogeneous, with spatially separated regions exhibiting markedly different mobilities~\cite{Ediger2000annurevphyschem,Kob1997prl,yamamoto1998prl,lacevic2003jcp,berthier2005science,dalle-ferrier2007pre,berthier2011revmodphys}.
The origin of such dynamic inhomogeneities remains a central and unsolved problem in glass physics. Various studies have explored possible structural signatures governing dynamical heterogeneity~\cite{kawasaki2007prl,tanaka2010natmat,kawasaki2011jpcm,leocmach2012natcommun,hirata2011natmat,tong2018prx,tong2019natcommun}.
However, the proposed descriptors are often system dependent, and no general structural measure has yet been established that quantitatively predicts dynamics across different models and conditions.

In recent years, machine learning (ML) has emerged as a powerful tool for uncovering structure--dynamics relationships in glass-forming systems by extracting relevant features directly from particle configurations.
Beyond handcrafted order parameters, ML-based descriptors—ranging from the concept of ``softness''~\cite{schoenholz2016natphys,schoenholz2017pnas} to recent deep learning models~\cite{boattini2020natcommun,boattini2021prl,coslovich2022jcp,jung2023prl,oyama2023font,liu2024jap,jung2024prb,roadmap2025natrevphys}—have demonstrated that static structure encodes significant information about future particle motion.
Among these approaches, graph neural networks (GNNs) have been particularly successful in this regard~\cite{bapst2020natphys}, outperforming traditional architectures by representing local environments through node and edge features. Recent developments in GNNs~\cite{shiba2023jcp,jiang2023jcp,pezzicoli2024scipost} have further refined this learnable mapping, providing a robust foundation for predicting particle-level dynamics directly from static snapshots. 
Collectively, these studies establish structure--dynamics prediction as a learnable mapping problem, bypassing the constraints of predefined physical descriptors.

Despite this progress, important conceptual issues remain unresolved.
In many existing studies, the temperature of the input data is fixed~\cite{bapst2020natphys,coslovich2022jcp,jung2023prl,shiba2023jcp,jiang2023jcp,pezzicoli2024scipost} or explicitly provided to the model~\cite{jung2024prb}.
As a result, many models are effectively trained and evaluated within a single temperature condition, leaving the relation between structure and dynamics across different temperatures largely unexplored.
This raises a fundamental question: do ML models genuinely learn a structure-dynamics mapping without temperature information, or do they exploit temperature information -- explicitly or implicitly -- to condition their predictions?
From a complementary perspective, one may instead ask how dynamical properties at different temperatures are related through static structure.
These questions are directly connected to physical interpretability and transferability -- the ability of a model to generalize to conditions beyond its training set.
A model that truly captures the structure-dynamics relation should therefore exhibit such generalization.
While recent works have explored transferability across systems or timescales~\cite{bapst2020natphys,jung2024prb,roadmap2025natrevphys}, generalization along the temperature axis -- without explicit temperature input -- remains largely unexplored, and it is unclear to what extent dynamic heterogeneities can be inferred from structure alone.

In this paper, we address this issue by performing supervised learning with a GNN trained on mixed-temperature datasets, without providing temperature as an explicit input. 
The model is designed to infer both particle-level dynamics and macroscopic temperature solely from static structure.
We then examine whether such a model can reproduce dynamic heterogeneity—including spatially correlated particle displacements and four-point correlation functions—from structure alone.
This enables us to assess if structural information supports interpolation in dynamics across temperatures.

This paper is organized as follows.
In Sec.~\ref{sec:dataset_method}, we describe the dataset details and machine learning methodology.
We show the prediction results of machine learning in Sec.~\ref{sec:result}.
We devote Sec.~\ref{sec:summary} to discussion and summary.

\section{Dataset details and machine learning methodology}
\label{sec:dataset_method}

\subsection{Dataset details}
\label{subsec:dataset_details}

\begin{figure}[tb]
    \centering
    \includegraphics[width=0.47\textwidth]{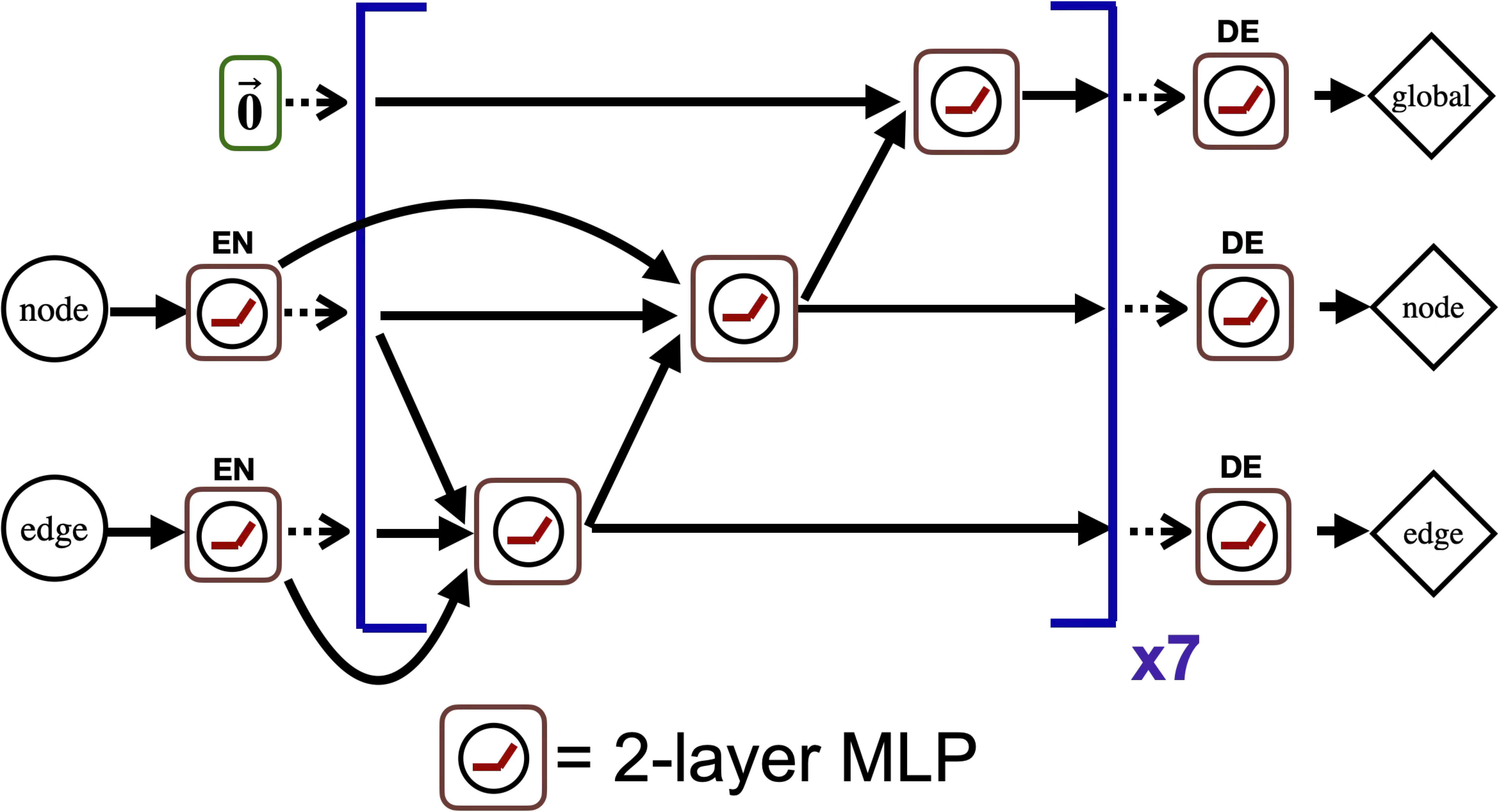}
    \caption{Schematic representation of the T-BOTAN architecture.
    Here, ``EN'', ``DE'', and ``MLP'' denote the encoder, decoder, and multilayer perceptron, respectively.
    Node features encode particle species, while edge features consist of the relative particle coordinates in the inherent structure and the interparticle distances in the thermal structure. Message passing iteratively updates node, edge, and global features, followed by node-, edge-, and global-level output heads that predict bond-breaking correlations, interparticle distance changes in the thermal structure, and temperature.
    }
    \label{fig:TBOTAN}
\end{figure}

\begin{table}[b]
\caption{\label{tab:table1} Temperature $T$ dependence of $\tau_{\alpha}$ in the MD simulation dataset.}
\begin{ruledtabular}
\begin{tabular}{c c c c c c c}
$T$ & $0.44$ & $0.47$ & $0.50$ & $0.56$ & $0.64$ & $0.80$ \\
$\tau_{\alpha}$ & $4120$ & $616$ & $217$ & $52$ & $16.5$ & $3.42$ \\
\end{tabular}
\end{ruledtabular}
\end{table}

In this study, we consider a three-dimensional Kob-Andersen binary Lennard-Jones (KALJ) mixture~\cite{kob1995pre1,kob1995pre2}.
The interaction potential is a linearly smoothed Lennard-Jones potential~\cite{shimada2018pre,toxvaerd2011jcp}, given by
\begin{equation}
    U(r)=u(r)-u(r_{\mathrm{c}})-(r-r_{\mathrm{c}})\left.\frac{du(r)}{dr}\right|_{r=r_{\mathrm{c}}},
\end{equation}
with
\begin{equation}
    u(r)=4\epsilon_{\mu\nu}\left[\left(\frac{\sigma_{\mu\nu}}{r}\right)^{12}-\left(\frac{\sigma_{\mu\nu}}{r}\right)^{6}\right],
\end{equation}
where $\mu,\nu\in\{\mathrm{A},\mathrm{B}\}$ denote particle species, $\epsilon_{\mu\nu}$ is the interaction energy scale, and $r_{\mathrm{c}}=2.5\sigma_{\mu\nu}$ is the cutoff distance.
This smoothing ensures continuity of both the potential and the force at $r=r_{\mathrm{c}}$.

The system consists of $N=4096$ particles, with $80\%$ being type A.
The KALJ mixture is non-additive, with parameters
$\epsilon_{\mathrm{AA}}=1.0$, $\epsilon_{\mathrm{AB}}=1.5$, $\epsilon_{\mathrm{BB}}=0.5$,
$\sigma_{\mathrm{AA}}=1.0$, $\sigma_{\mathrm{AB}}=0.8$, and $\sigma_{\mathrm{BB}}=0.88$.
Accordingly, units of length, energy, and time are set to $\sigma_{\mathrm{AA}}$, $\epsilon_{\mathrm{AA}}$, and $\sqrt{m\sigma_{\mathrm{AA}}^2/\epsilon_{\mathrm{AA}}}$, respectively.
The number density is fixed at $\rho=N/V=1.2$.
Full details of the dataset generation are provided in Ref.~\cite{shiba2023jcp}.

Our model utilizes a hybrid input set consisting of instantaneous thermal structures at $t=0$ -- prepared through long-time relaxation at each temperature--  and their corresponding inherent structures, which are obtained by quenching the $t=0$ thermal structures. 
This inclusion of inherent structures is motivated by its ability to offer a definitive structural fingerprint of glassy dynamics~\cite{jung2024prb,pezzicoli2024scipost,roadmap2025natrevphys}, whereas the instantaneous thermal structures retain the full information of the systems state. 
Therefore, this combination ensures a robust representation of the local environment surrounding each particle.
Inherent structures characterize the basins of the potential-energy landscape, providing a geometric reference for the subsequent dynamics. 

To obtain inherent structures, instantaneous thermal structures at $t=0$ are quenched using the FIRE algorithm~\cite{Bitzek2006prl,Shuang2019compmatsci}.
We adopt the same parameter settings as in Ref.~\cite{Bitzek2006prl}, except for the initial and maximum time steps, which are set to $dt_{\mathrm{init}}=0.002$ and $dt_{\mathrm{max}}=0.02$, following Ref.~\cite{Alkemade2023jcp}.
A configuration is considered mechanically equilibrated when the average force magnitude per particle is smaller than $10^{-12}\epsilon_{\mathrm{AA}}/\sigma_{\mathrm{AA}}$.

In this study, the particle propensity, defined as the isoconfigurational average of the particle mobility~\cite{widmercooper2004prl,widmercooper2008natphys}, at time $t=\tau_{\alpha}$ is obtained from 32 independent trajectories.
For each temperature, equilibrium configurations are first prepared by MD simulations as described above.
Starting from each configuration, multiple runs are generated by assigning different sets of initial velocities drawn from the Maxwell-Boltzmann distribution at the target temperature while keeping the particle positions identical.
These trajectories are evolved up to time $t=\tau_{\alpha}$, where $\tau_{\alpha}$ is defined by $F_s(q,\tau_{\alpha})=1/e$ for type-A particles (Table~\ref{tab:table1}).
\subsection{GNN model architecture}
Our GNN model, hereafter referred to as T-BOTAN, is developed as a direct extension of the BOnd TArgeting Network (BOTAN) proposed by Shiba \textit{et al.}~\cite{shiba2023jcp}. 
Following the framework of BOTAN and the work by Bapst \textit{et al.}~\cite{bapst2020natphys}, T-BOTAN is based on the interaction network architecture~\cite{battaglia2018}. In this framework, the latent features of nodes and edges within a graph are mutually computed through message-passing steps.
These steps involve a pair of two-layer multilayer perceptrons (MLPs), which update the feature vectors for nodes, 
$h_\nu^m$ and edges, $h_{e_{\nu,u}}^m$, respectively. 
Hereafter, we adhere to the notation used in Ref.~\cite{shiba2023jcp}.

The architecture is illustrated in Fig.~\ref{fig:TBOTAN}. 
To relate the local dynamics to the macroscopic information, we introduce an additional MLP block for a global feature vector $h_{\gamma}^m$. 
In our model, we do not employ a separate global encoder; instead, the global feature vector is initialized as a 64-dimensional zero vector, $h_\gamma^0 = {\bf 0}$. During the message-passing process, the global feature is updated as follows:
\begin{equation}
h_\gamma^m = {\rm MLP} \left( h_\gamma^{m-1} \oplus \frac{1}{N} \sum_{\nu \in V} h_\nu^{m} \right),
\end{equation}
where $m$ stands for the iteration index of $n$ repeat cycles ($m = 1,\ldots n$).
Here $\oplus$ represents the concatenation operator, and the summation is performed over all nodes $V$ in the graph to aggregate local information into the global context.
Note that the global feature is not fed back to the node or edge updates during message passing. 
It is used solely for aggregating system-level information and producing the final output.
The number of repeat cycles is set to $n=7$ consistent with BOTAN and related studies~\cite{bapst2020natphys,shiba2023jcp}. 
Finally, the updated global feature is passed to the decoder MLP to produce the final output:
\begin{equation}
h_\gamma^{\rm out} = {\rm DE} (h_\gamma^n ).
\end{equation}
The MLPs and decoders added in this paper, as well as the others, have two hidden layers, where each consists of 64 units using the rectified linear units (ReLU) as the activation function. 

\subsection{Machine learning procedure}
Input particle configurations are represented as graphs.
Node features correspond to particle species, while edge features consist of relative particle coordinates in the inherent structure and the interparticle distances in the instantaneous thermal structure.
Edges are constructed by connecting particles within a cutoff distance of $2\sigma_{\mathrm{AA}}$ in the inherent structure.

The network predicts three types of outputs: 
(i) node-level bond-breaking correlation functions $\{C_{\mathrm{B}}^j(t)\}$~\cite{Guiselin2022natphys}, 
(ii) edge-level changes in interparticle distances $\mathcal{E}_{jk}=r_{jk}(t)-r_{jk}(0)$ in the instantaneous thermal structure, 
and (iii) the global temperature $T_{\rm{pred}}$.

The bond-breaking correlation function for particle $j$ is defined as
\begin{equation}
    C_{\mathrm{B}}^j(t)=\left\langle \frac{n_j(t|0)}{n_j} \right\rangle_{\mathrm{iso}},
\end{equation}
where $n_j$ is the number of nearest neighbors satisfying $r_{jk}<1.3\sigma_{jk}$ at $t=0$, and $n_j(t|0)$ is the number of those neighbors that satisfy $r_{jk}<1.6\sigma_{jk}$ at time $t$.
The angular brackets $\langle\cdots\rangle_{\mathrm{iso}}$ denote the isoconfigurational ensemble average introduced in Sec.~\ref{subsec:dataset_details}.

The total loss function is defined as
\begin{equation}
    \mathcal{L}_{\mathrm{all}}
    =\lambda_{\RM{n}}\mathcal{L}_{\mathrm{node}}
    +\lambda_{\RM{e}}\mathcal{L}_{\mathrm{edge}}
    +\lambda_{\RM{g}}\mathcal{L}_{\mathrm{global}},
\end{equation}
where $\mathcal{L}_{\mathrm{node}}$, $\mathcal{L}_{\mathrm{edge}}$, and $\mathcal{L}_{\mathrm{global}}$ are the loss terms for node, edge, and global outputs, respectively.
We set $\lambda_{\RM{n}}=\lambda_{\RM{e}}=5\times10^{-1}$ as suggested in Ref.~\cite{shiba2023jcp} and $\lambda_{\RM{g}}=10^{-2}$.
The robustness of the results with respect to $\lambda_{\RM{g}}$ is verified in Appendix~\ref{sec:S3}.

To accurately reproduce dynamic heterogeneity, the node loss is defined as
\begin{equation}
    \mathcal{L}_{\mathrm{node}}
    =\mathcal{L}_{\mathrm{MSE}}
    +w_{\mathrm{v}}\mathcal{L}_{\mathrm{Var}}
    +w_{\mathrm{c}}\mathcal{L}_{\mathrm{Corr}},
\end{equation}
with
\begin{equation}
    \begin{cases}
        \displaystyle
        \mathcal{L}_{\mathrm{MSE}}
        =\frac{1}{N}\sum_j
        \frac{(C_{\mathrm{B,pred}}^j-C_{\mathrm{B,MD}}^j)^2}
        {\mathrm{Var}(\{C_{\mathrm{B,MD}}\})},\\[6pt]
        \displaystyle
        \mathcal{L}_{\mathrm{Var}}
        =\frac{\big|\mathrm{Var}(\{C_{\mathrm{B,pred}}\})
        -\mathrm{Var}(\{C_{\mathrm{B,MD}}\})\big|}
        {\mathrm{Var}(\{C_{\mathrm{B,MD}}\})},\\[6pt]
        \displaystyle
        \mathcal{L}_{\mathrm{Corr}}
        =\sum_{d=2,4,6}
        \frac{\big|F(\{C_{\mathrm{B,pred}}\},d)
        -F(\{C_{\mathrm{B,MD}}\},d)\big|}
        {F_{\mathrm{N}}}
    \end{cases}
\end{equation}
as proposed by Jung \textit{et al.}{~\cite{jung2023prl,jung2024prb}, where $\RM{Var}(\{\mathcal{X}\})$ is the variance of $\{\mathcal{X}\}$.
Here,
\begin{equation}
    \begin{cases}
        \displaystyle
        F(\{\mathcal{X}\},d)
        =\frac{\sum_{j,k}\delta\mathcal{X}^j\delta\mathcal{X}^k
        e^{-(r_{jk}-d)^2/2}}
        {\sum_j(\delta\mathcal{X}^j)^2},\\[6pt]
        F_{\mathrm{N}}
        =\big|F(\{C_{\mathrm{B,MD}}\},d_{\mathrm{min}})
        -F(\{C_{\mathrm{B,MD}}\},d_{\mathrm{max}})\big|
    \end{cases}
\end{equation}
with $\delta\mathcal{X}^j=\mathcal{X}^j-\bar{\mathcal{X}}$.

$\mathcal{L}_{\RM{edge}}$ and $\mathcal{L}_{\RM{global}}$ are mean-squared errors, where $\mathcal{L}_{\mathrm{edge}}$ is normalized by $\mathrm{Var}(\{C_{\mathrm{B,MD}}\})$.
To stabilize training, we set $w_{\mathrm{v}}=w_{\mathrm{c}}=0$ for the first 1000 epochs, where $\mathcal{L}_{\RM{MSE}}$ is sufficiently stable, and subsequently switch to $w_{\mathrm{v}}=1$ and $w_{\mathrm{c}}=5\times10^{-1}$ as suggested in Ref.~\cite{jung2023prl}.
In Fig.~\ref{fig:S_loss_all} of Appendix~\ref{sec:S1}, we present the dependence of each loss function on the number of epochs.

In this study, a single T-BOTAN model is trained on a mixed-temperature dataset consisting of configurations at $T=0.44$, $0.47$, $0.64$, and $0.80$, evaluated at $t=\tau_{\alpha}$.
For each temperature, 100 configurations are used for training.
Importantly, configurations at different temperatures are combined during training, and no temperature information is provided as an input feature.
To evaluate generalization, separate test datasets are prepared for both trained and untrained temperatures.
Configurations at $T=0.50$ and $0.56$ are used exclusively for testing and are never seen during training.
Test configurations at trained temperatures are also disjoint from the training set.
All results are averaged over five independent trainings with different random seeds.

\section{Results}
\label{sec:result}
\begin{figure*}[tb]
    \centering
    \includegraphics[width=0.95\textwidth]{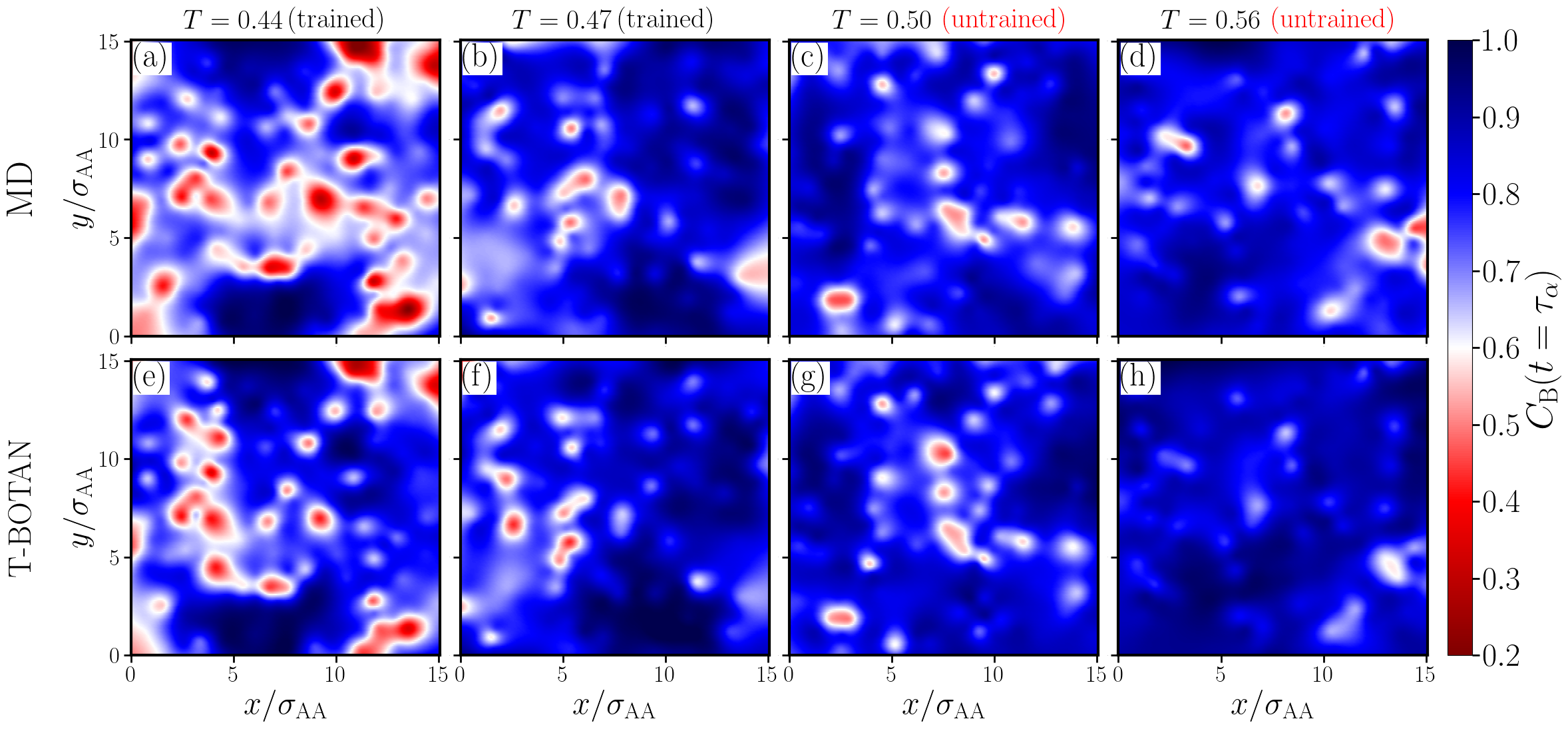}
    \caption{Comparison of MD simulation data ((a)-(d)) and T-BOTAN predictions ((e)-(h)) for the spatial distribution of $C_{\RM{B}}(t=\tau_{\alpha})$ at $T=0.44, 0.47, 0.50$, and $0.56$.
    All data are plotted for the same three-dimensional particle configuration, in the cross section $11.1<z<11.9$.}
    \label{fig:CB}
\end{figure*}
Figure~\ref{fig:CB} compares representative snapshots of the spatial distribution of $C_{\RM{B}}(t=\tau_{\alpha})$ obtained from molecular dynamics (MD) simulations and from T-BOTAN predictions.
For the temperatures included in the training set, $T=0.44$ and $0.47$, T-BOTAN accurately reproduces the strongly heterogeneous spatial patterns observed in the MD data.
Regions of enhanced and suppressed mobility, as well as their spatial organization, are well captured despite the fact that temperature is not provided as an explicit input feature.
This agreement demonstrates that the GNN successfully learns structure-dependent particle dynamics directly from static structural information.
We also present the results of $C_{\RM{B}}(t=\tau_{\alpha})$ for $T=0.64$ and $0.80$ in Fig.~\ref{fig:S_CB} of Appendix~\ref{sec:S2}.
More strikingly, for the \textit{untrained} temperatures ($T=0.50$ and $0.56$), the predicted spatial patterns exhibit a qualitative agreement with the MD results.
As shown in Figs.~\ref{fig:CB}(g) and (h), the network correctly identifies specific regions of high and low mobility, preserving the spatial topology of dynamic heterogeneity.
Although the precise contrast is slightly biased toward the nearest trained temperatures, the essential structural motifs governing local facilitation and arrest are robustly detected.
This tendency for predictions to be biased toward the nearest trained temperatures is also reflected in the Pearson correlation coefficients between the predicted and MD values of $C_{\RM{B}}^j$, shown in Fig.~\ref{fig:S_pearson} of Appendix~\ref{sec:S1}.
This successful generalization to untrained temperatures provides evidence that the structure-dynamics relationship learned by T-BOTAN is not a mere memorization of state points, but a transferable physical mapping rooted in local structural instability.

\begin{figure}[tb]
    \centering
    \includegraphics[width=0.47\textwidth]{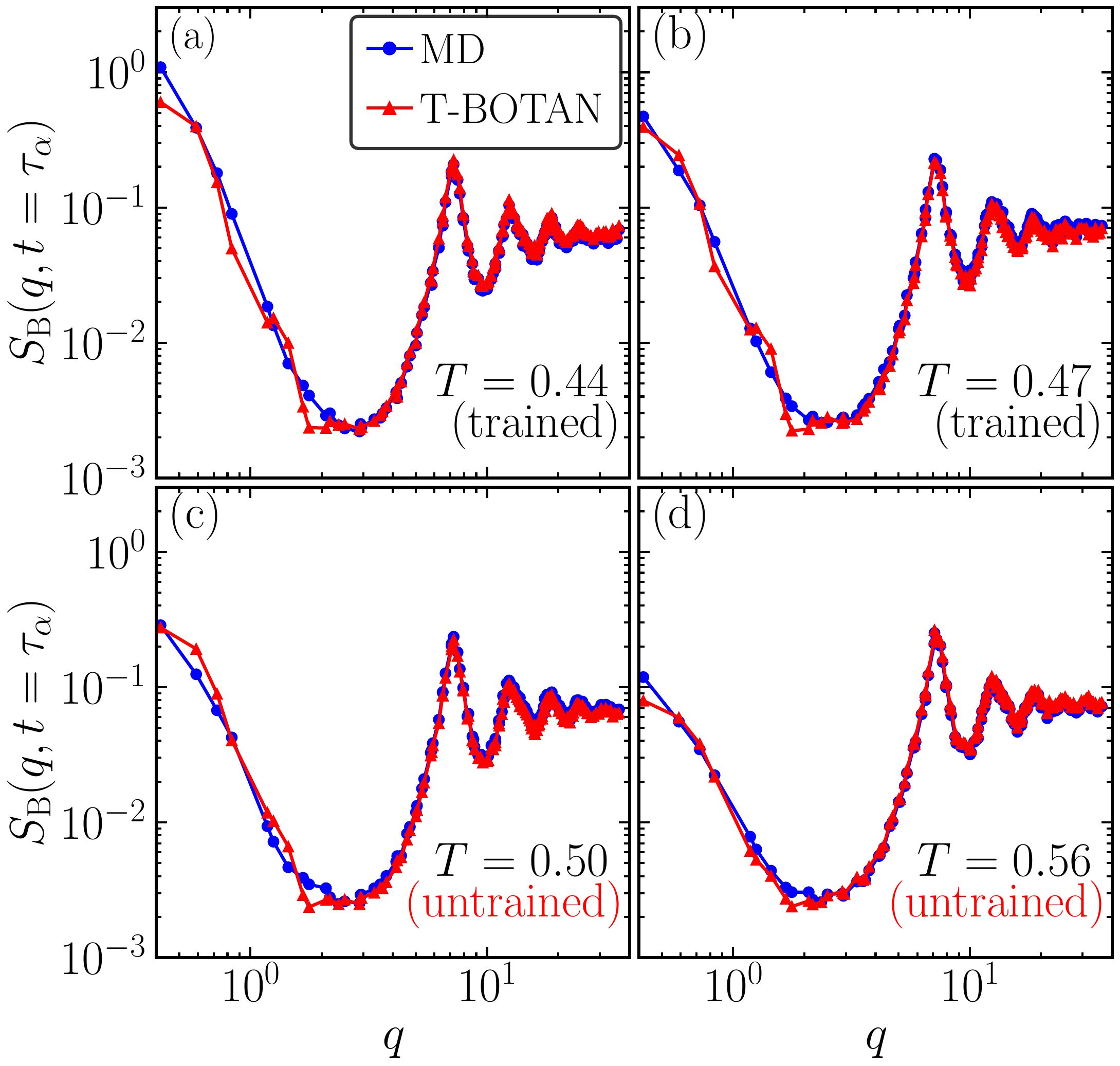}
    \caption{Comparison of $S_{\RM{B}}(q,t=\tau_{\alpha})$ from MD simulation data with the prediction of T-BOTAN for various $T$. 
    (a) $T=0.44$ (b) $T=0.47$ (c) $T=0.50$ (d) $T=0.56$}
    \label{fig:SB}
\end{figure}
To quantify the spatial correlations underlying these heterogeneous dynamics, we analyze the four-point dynamic structure factor $S_{\RM{B}}(q,t)$,
\begin{equation}
    S_{\RM{B}}(q,t)=\frac{1}{N_{\RM{A}}}\left\langle W(\bm{q},t)W(-\bm{q},t) \right\rangle,
    \label{eq:SB}
\end{equation}
where $W(\bm{q},t)=\sum_{j\in N_{\RM{A}}}[C_{\RM{B}}^j(t)-\bar{C}_{\RM{B}}(t)]e^{i\bm{q}\cdot\bm{r}_j(0)}$~\cite{yamamoto1998prl,shiba2012pre}.
Herem, the T-BOTAN prediction of $S_{\RM{B}}(q,t)$ is calculated by substituting the predicted node-level correlations $\{C_{\RM{B}}^j\}$ into Eq.~(\ref{eq:SB}).
Figure~\ref{fig:SB} shows $S_{\RM{B}}(q,t=\tau_{\alpha})$ obtained from MD simulations and T-BOTAN predictions for trained ($T=0.44,0.47$) and untrained temperatures ($T=0.50,0.56$).
As temperature decreases, the MD results exhibit a pronounced enhancement of $S_{\RM{B}}(q)$ at small wave numbers, signaling the growth of a dynamic correlation length associated with cooperative particle motion.
For the trained temperatures $T=0.44$ and $0.47$, T-BOTAN reproduces the overall wave-number dependence of $S_{\RM{B}}(q)$ over the full $q$ range.
We also present the results of $S_{\RM{B}}(q,t=\tau_{\alpha})$ for $T=0.64$ and $0.80$ in Fig.~\ref{fig:S_SB} of Appendix~\ref{sec:S2}.
This demonstrates that information about the spatial extent of dynamic heterogeneity is successfully encoded in the structural representation learned by the network, even without explicit temperature input.
Crucially, for the untrained temperatures $T=0.50$ and $0.56$, the predicted $S_{\RM{B}}(q)$ remains in good agreement with the MD data.
In particular, the enhancement of $S_{\RM{B}}(q)$ at small wave numbers, which reflects the growth of the dynamic correlation length, is quantitatively reproduced.
This suggests that the network captures the intrinsic structural signatures governing collective dynamics even at untrained temperatures.
While the slight deviation of the predictions toward nearby trained temperatures reflects the discrete nature of the training manifold, the successful capture of the growing correlation length underscores a robust generalization capability.
This indicates that the dominant length scales of dynamic heterogeneity are encoded in the static structure in a physically transferable manner.
In this sense, T-BOTAN captures the systematic evolution of cooperative dynamics with temperature at the level of physically relevant observables, even when the temperature itself is not explicitly learned as a continuous variable.

\begin{figure}[tb]
    \centering
    \includegraphics[width=0.35\textwidth]{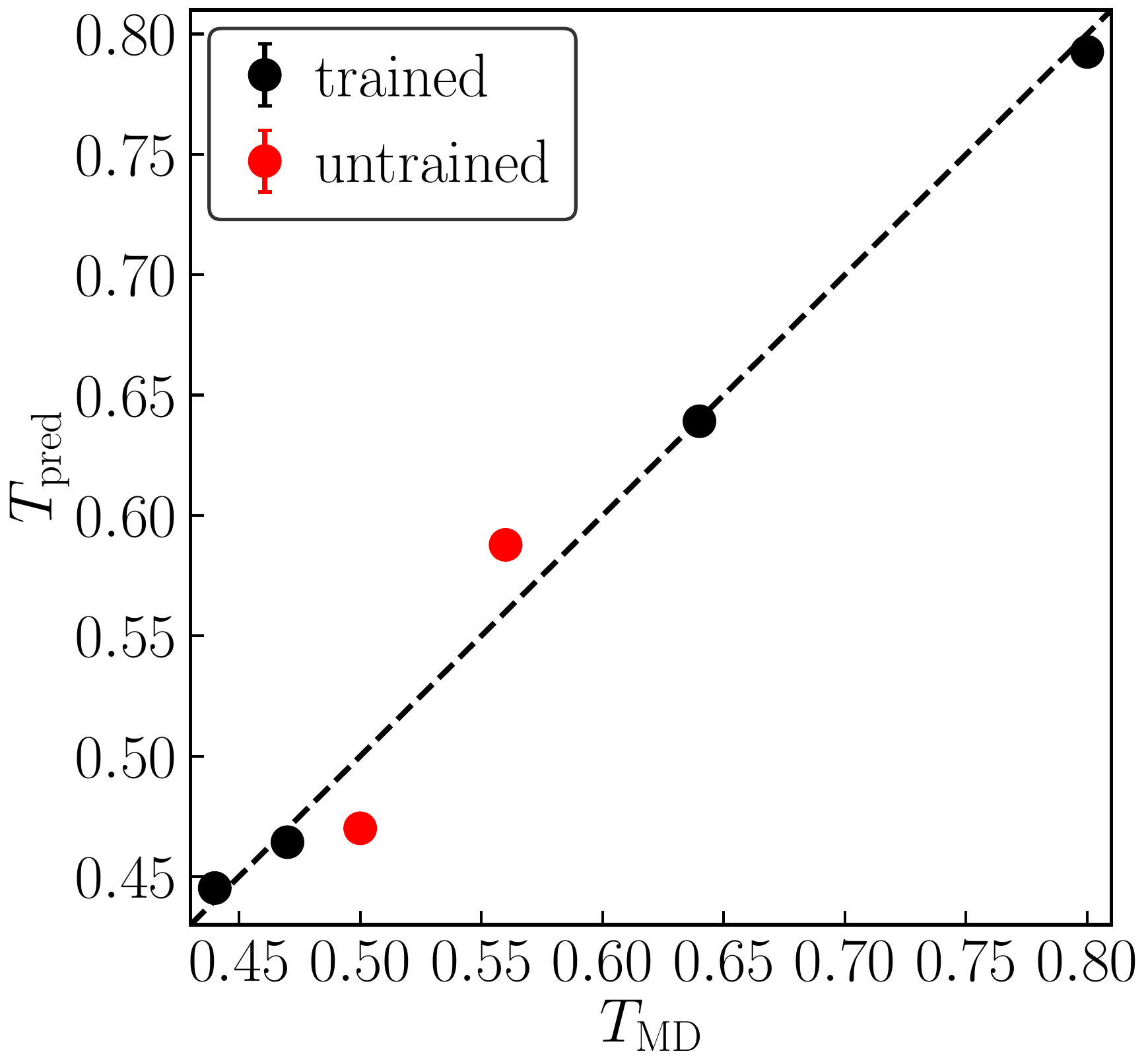}
    \caption{Comparison of $T_{\RM{pred}}$ with $T_{\RM{MD}}$. Black symbols denote temperatures used as training data, and red symbols denote temperatures not used for training. 
    Error bars represent the standard deviation over different samples, but are smaller than the symbol size and thus not visible.
    The dashed line indicates $T_{\RM{pred}}=T_{\RM{MD}}$.}
    \label{fig:Tpred_vs_TMD}
\end{figure}
As we have shown above, T-BOTAN can accurately predict dynamic heterogeneity even when trained on mixed-temperature data without explicit temperature labels.
This suggests that the network may implicitly capture structural signatures associated with temperature.
To examine this possibility, we next analyze whether the model can infer the temperature of a configuration directly from structural information.

Figure~\ref{fig:Tpred_vs_TMD} compares the temperature predicted by T-BOTAN, $T_{\RM{pred}}$, with the actual simulation temperature, $T_{\RM{MD}}$.
For temperatures included in the training set ($T=0.44, 0.47, 0.64$, and $0.80$), $T_{\RM{pred}}$ agrees well with $T_{\RM{MD}}$.
This demonstrates that T-BOTAN can identify temperature-dependent structural signatures without explicit temperature input.
More generally, predicting $T_{\RM{MD}}$ from structural configurations can be viewed as a statistical inference of temperature based solely on geometric information.
In our model, the edge features incorporate relative coordinates from the inherent structure and interparticle distances from the instantaneous thermal structure.
Accordingly, the predicted temperature should be interpreted as a data-driven estimate derived from structural descriptors, rather than a direct evaluation of a thermodynamic quantity.
Importantly, the network does not access forces, energies, or any explicit thermodynamic variables.
Instead, it learns temperature-dependent patterns embedded in particle configurations.
This robustness highlights that structural information alone is sufficient to encode thermodynamic properties.
These observations are conceptually related to the recent work by Janzen \textit{et al.}~\cite{janzen2024prm}, which demonstrated that structural descriptors such as the radial distribution function of glassy systems can encode information about thermal history and aging state.
In a similar spirit, our results indicate that structural configurations retain statistically detectable correlations with their preparation temperature, which can be recovered through data-driven inference.

To clarify the respective roles of the two structural representations, we additionally examined models trained using only inherent structures or only instantaneous thermal structures as inputs (see Fig.~\ref{fig:S_Tpred_vs_TMD_IS_vs_TS} of Appendix~\ref{sec:S4}).
Using inherent structures alone, the predicted temperature corresponds to the equilibrium temperature at which the configurations were prepared.
This observation is consistent with the potential energy landscape framework, in which inherent structures characterize basins of the energy landscape and encode thermodynamic information about the parent equilibrium state~\cite{stillinger1982pra,stillinger1995science,sastry1998nat,heuer2008jpcm}.
In contrast, models trained solely on instantaneous thermal structures yield predictions that reflect the thermodynamic temperature associated with the equilibrated instantaneous structures.
In both cases, the temperature is obtained as a statistical estimate from structural information, and the overall predictive behavior remains qualitatively similar.

For temperatures not included in the training set ($T=0.50$ and $0.56$), which lie between the trained temperatures, $T_{\RM{pred}}$ remains close to $T_{\RM{MD}}$, demonstrating the robustness of the learned structural representation beyond the training data.
Although the predictions do not interpolate perfectly between the trained values, they systematically reflect the structural proximity to neighboring trained temperatures.
This tendency is qualitatively consistent with the predictions of particle dynamics shown in Figs.~\ref{fig:CB} and \ref{fig:SB}.
Additionally, we present the distribution function of $T_{\RM{pred}}/T_{\RM{MD}}$ in Fig.~\ref{fig:S_Tpred_dist} of Appendix~\ref{sec:S2}.

\begin{figure}[tb]
    \centering
    \includegraphics[width=0.47\textwidth]{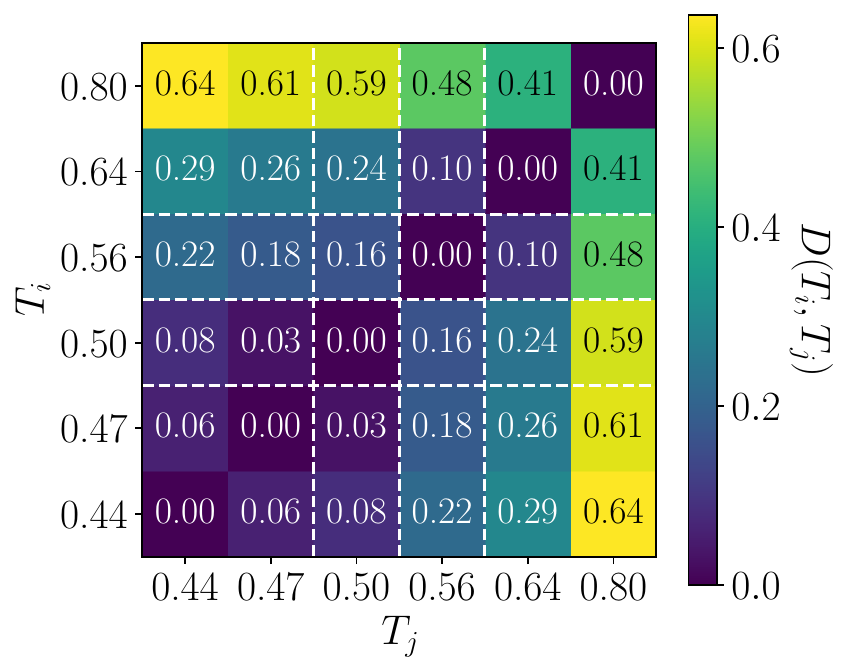}
    \caption{Pairwise distances $D(T_i,T_j)$ between mean global latent representations learned by T-BOTAN at different temperatures.
    For each temperature, global embeddings obtained at the final message-passing layer (before decoding) are averaged over configurations and random seeds. 
    Small distances indicate that the network encodes configurations at the two temperatures in a similar manner, revealing that temperatures form discrete clusters in the learned feature space.
    The diagonal components are by definition zero.}
    \label{fig:embedding_distance}
\end{figure}
To elucidate the origin of this discrete temperature prediction, Fig.~\ref{fig:embedding_distance} examines how the network internally represents particle configurations at different temperatures.
For each input configuration, the trained GNN produces a fixed-length (in this study 64-dimensional) vector associated with the global representation at the output of the final message-passing layer, before the decoding stage that generates node-, edge-, and global-level predictions.
This vector provides a compact summary of the structural features extracted by the network at the system level.
We denote this internal global representation obtained from configuration $c$ with random seed $s$ as $\bm{u}_{s,c}$.
Thus, $\bm{u}_{s,c}$ represents the position of a configuration in the network’s learned global feature space prior to temperature prediction.
For each temperature $T$, we average these latent representations over all configurations in the test set and over different random seeds, and define the mean embedding as
\begin{equation}
    \langle \bm{u}(T) \rangle
    = \frac{1}{N_{\RM{seed}} N_{\RM{conf}}}
    \sum_{s=1}^{N_{\RM{seed}}}
    \sum_{c=1}^{N_{\RM{conf}}}
    \bm{u}_{s,c}(T),
\end{equation}
where $N_{\RM{seed}}$ and $N_{\RM{conf}}$ denote the number of random seeds and the number of configurations, respectively.
We then quantify the similarity between two temperatures $T_i$ and $T_j$ by computing the Euclidean distance between their mean representations,
\begin{equation}
    D(T_i,T_j)
    = \left| \langle\bm{u}(T_i)\rangle - \langle\bm{u}(T_j)\rangle \right|.
\end{equation}
A small value of $D(T_i,T_j)$ indicates that the network encodes configurations at the two temperatures in a similar manner, whereas a large value indicates distinct internal representations.
Figure~\ref{fig:embedding_distance} shows that the distance between embeddings at $T=0.47$ and $T=0.50$ is extremely small, indicating that the network regards these structures as highly similar.
A comparable behavior is observed between $T=0.56$ and $T=0.64$.
These results demonstrate that, within the learned representation, temperature is organized into discrete clusters associated with the trained temperatures.
Consequently, configurations at untrained temperatures are projected onto the nearest learned cluster rather than being placed continuously along a temperature axis.

Taken together, the ability of T-BOTAN to make accurate interpolative predictions without explicit temperature input demonstrates that a robust structure-dynamics mapping is maintained across different thermodynamic states.
Notably, the reproduction of dynamic heterogeneity reveals that this essential relationship transcends specific temperatures.
These findings support the view that static particle configurations encode not only local particle structures responsible for dynamical heterogeneity with medium correlation length scale but also macroscopic thermodynamic information.
This indicates that both collective dynamics and thermodynamic state variables are implicitly organized within the geometry of configuration space itself.

\section{Discussion and Summary}
\label{sec:summary}
In this paper, we introduced a graph neural network, termed T-BOTAN, to predict particle-level dynamics and macroscopic temperature solely from static particle configurations.
Unlike most previous machine-learning studies of glassy dynamics, which either fix the temperature of the input data or provide temperature as an explicit input feature, our approach removes temperature information from the input.
This design allows us to directly test a fundamental physical question: to what extent are temperature-dependent dynamical properties intrinsically encoded in static structure?

Using a single T-BOTAN trained on mixed-temperature datasets, we demonstrated that dynamic heterogeneity—including the spatial distribution of the bond-breaking correlation function and four-point correlation functions—can be reproduced with good accuracy from static particle configurations.
Importantly, the predictions remain reasonably accurate not only at the trained temperatures but also at previously untrained temperatures lying within the trained temperature range.
In particular, the successful interpolation of four-point correlations without explicit temperature input indicates that the network does not rely on temperature labels as auxiliary variables, but instead extracts structural features that systematically evolve across thermodynamic states.

Recently, Jung \textit{et al.}~\cite{jung2024prb} showed that dynamic heterogeneity can be predicted when temperature is explicitly provided as an input to their model, demonstrating transferability across temperatures under supervised thermodynamic conditioning. 
In contrast, the present study removes this conditioning entirely. 
Our results therefore suggest a stronger statement: even in the absence of explicit thermodynamic information, static configurations contain sufficient structural information to interpolate temperature-dependent dynamic heterogeneity. 
To our knowledge, this provides the first explicit demonstration that four-point dynamical correlations can be interpolatively predicted across temperatures using structure alone.

In addition to predicting dynamical observables, the network can also estimate the macroscopic temperature from structural information.
Configurations at similar temperatures are embedded closely in the learned feature space, leading to reasonably accurate predictions within the trained temperature window.
Since neither temperature nor forces, energies, or other explicit thermodynamic variables are provided as inputs, this result indicates that the network identifies structural features that correlate with thermodynamic state.

Overall, these findings suggest that T-BOTAN interprets a glassy configuration primarily through structural characteristics.
A particularly notable result is that the model retains predictive accuracy even when the input configurations are inherent structures prepared at different parent temperatures.
Because inherent structures eliminate thermal fluctuations and thus reduce structural differences between temperatures, the persistence of predictive accuracy highlights the robustness of the structural information extracted by the network.
This observation suggests that important aspects of temperature-dependent glassy dynamics are encoded in the configurational organization.
These results are broadly consistent with theoretical approaches emphasizing configurational contributions to glassy slowing down~\cite{kawasaki2007prl,tanaka2010natmat,kawasaki2011jpcm,hirata2011natmat,royall2015physrepo,tong2018prx,tong2019natcommun}.
Whether this structural encoding remains predictive outside the trained temperature range or across qualitatively distinct dynamical regimes remains an important direction for future investigation.

\section*{acknowledgments}
H.B. would like to thank ``THERS Make New Standards Program for the Next Generation Researchers''.
T.K. acknowledges support from the JST FOREST Program (Grant No. JPMJFR212T), AMED Moonshot Program (Grant No. JP22zf0127009), JSPS KAKENHI (Grant No. JP24H02204), and Takeda Science Foundation.
H.S. acknowledges support from JSPS KAKENHI  (Grant Nos. 22K03546 and 25H01109). 
Computational resources werfe provided by the ``Joint Usage/Research Center for Interdisciplinary Large-scale Information Infrastructures (JHPCN)'' and ``High Performance Computing Infrastructure (HPCI)'' in Japan (project ID: jh250078).

\section*{data availability}
The datasets of instantaneous thermal configurations used in this study that were not covered in Ref.~\cite{shiba2023jcp} will be made publicly available upon publication.

\appendix

\section{Loss functions and training details}
\label{sec:S1}
\begin{figure}
    \centering
    \includegraphics[width=0.48\textwidth]{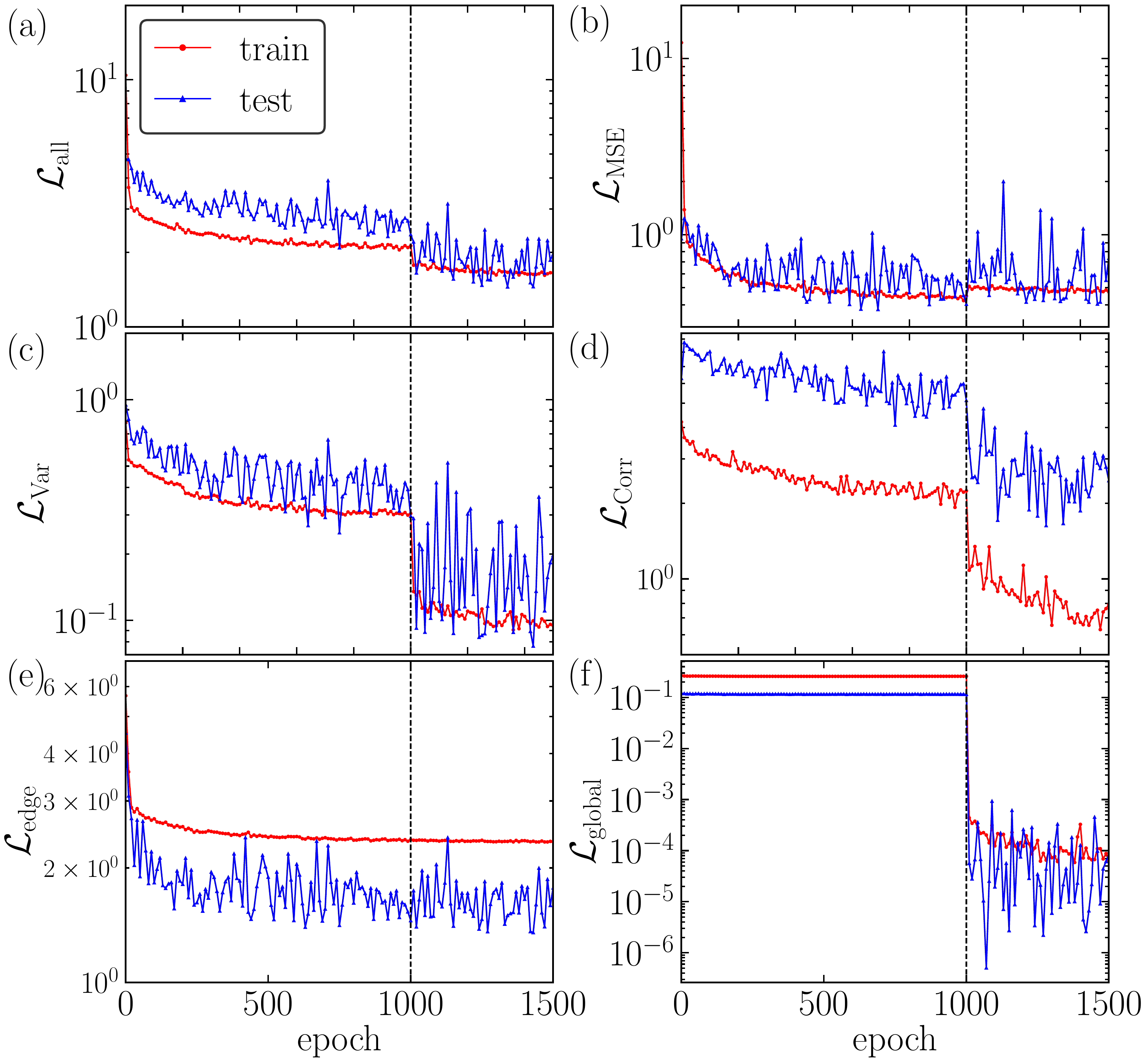}
    \caption{Training and test losses. 
    The training loss is averaged over the four temperatures used for training, while the validation loss is evaluated at the lowest training temperature $T=0.44$. 
    The vertical dashed line indicates the epoch at which the variance, spatial correlation, and global terms become activated.
    (a) $\mathcal{L}_{\RM{all}}$. (b) $\mathcal{L}_{\RM{MSE}}$. (c) $\mathcal{L}_{\RM{Var}}$. (d) $\mathcal{L}_{\RM{Corr}}$. (e) $\mathcal{L}_{\RM{edge}}$. (f) $\mathcal{L}_{\RM{global}}$.}
    \label{fig:S_loss_all}
\end{figure}
\begin{figure}
    \centering
    \includegraphics[width=0.48\textwidth]{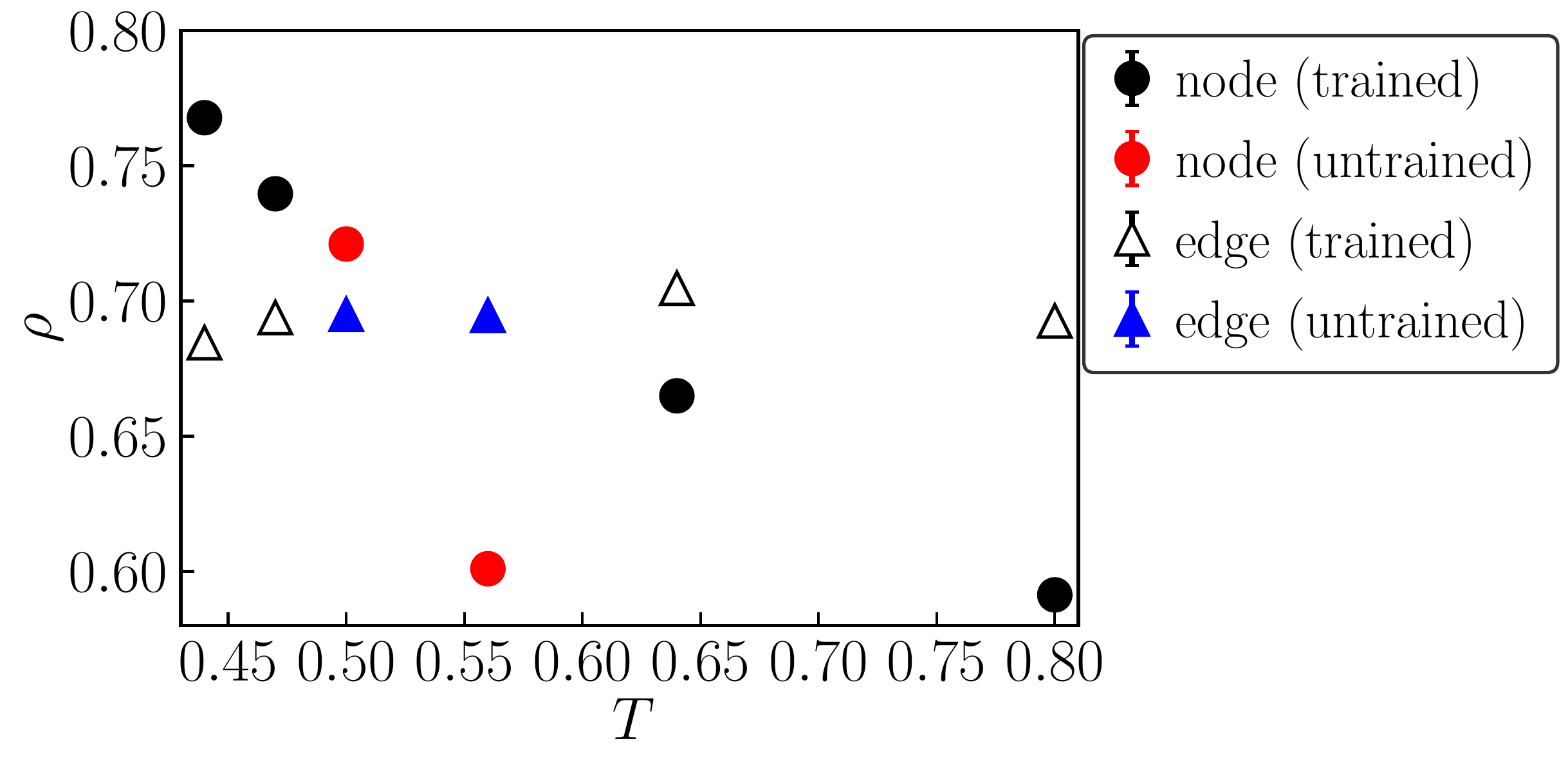}
    \caption{Pearson correlation coefficients $\rho$ between the T-BOTAN predictions and the MD simulation results for various $T$, evaluated for both node- and edge-level quantities.
    Filled symbols represent temperatures included in the training dataset, whereas open symbols denote temperatures not used for training.
    Error bars represent the standard deviation over different samples, but are smaller than the symbol size and thus not visible.
    }
    \label{fig:S_pearson}
\end{figure}
Figure~\ref{fig:S_loss_all} shows the loss functions for the training and test datasets.
Although the test loss exhibits relatively large fluctuations, both losses decrease as the number of epochs increases and eventually converge to approximately constant average values.

Figure~\ref{fig:S_pearson} shows the temperature dependence of the Pearson correlation coefficients $\rho$ for the node and edge outputs between the T-BOTAN predictions and the MD simulation results, defined as
\begin{equation}
    \rho = \frac{\sum_j \delta x^j \, \delta y^j}
    {\sqrt{\sum_j (\delta x^j)^2} \sqrt{\sum_j (\delta y^j)^2}},
\end{equation}
where $\delta \mathcal{X}^j = \mathcal{X}^j - \bar{\mathcal{X}}$.
The coefficient $\rho$ quantifies the strength and direction of the linear relationship between two variables.
As $T$ decreases, $\rho$ for the node outputs ($\{C_{\RM{B}}^j\}$) monotonically increases, even at $T = 0.50$, which is not included in the training dataset.
However, the value at $T = 0.56$ does not follow this monotonic trend.
For the edge outputs, $\rho$ is approximately independent of $T$, which is consistent with the results reported by Shiba~\textit{et al.}~\cite{shiba2023jcp}.

\section{Supplemental data}
\label{sec:S2}

\begin{figure}
    \centering
    \includegraphics[width=0.48\textwidth]{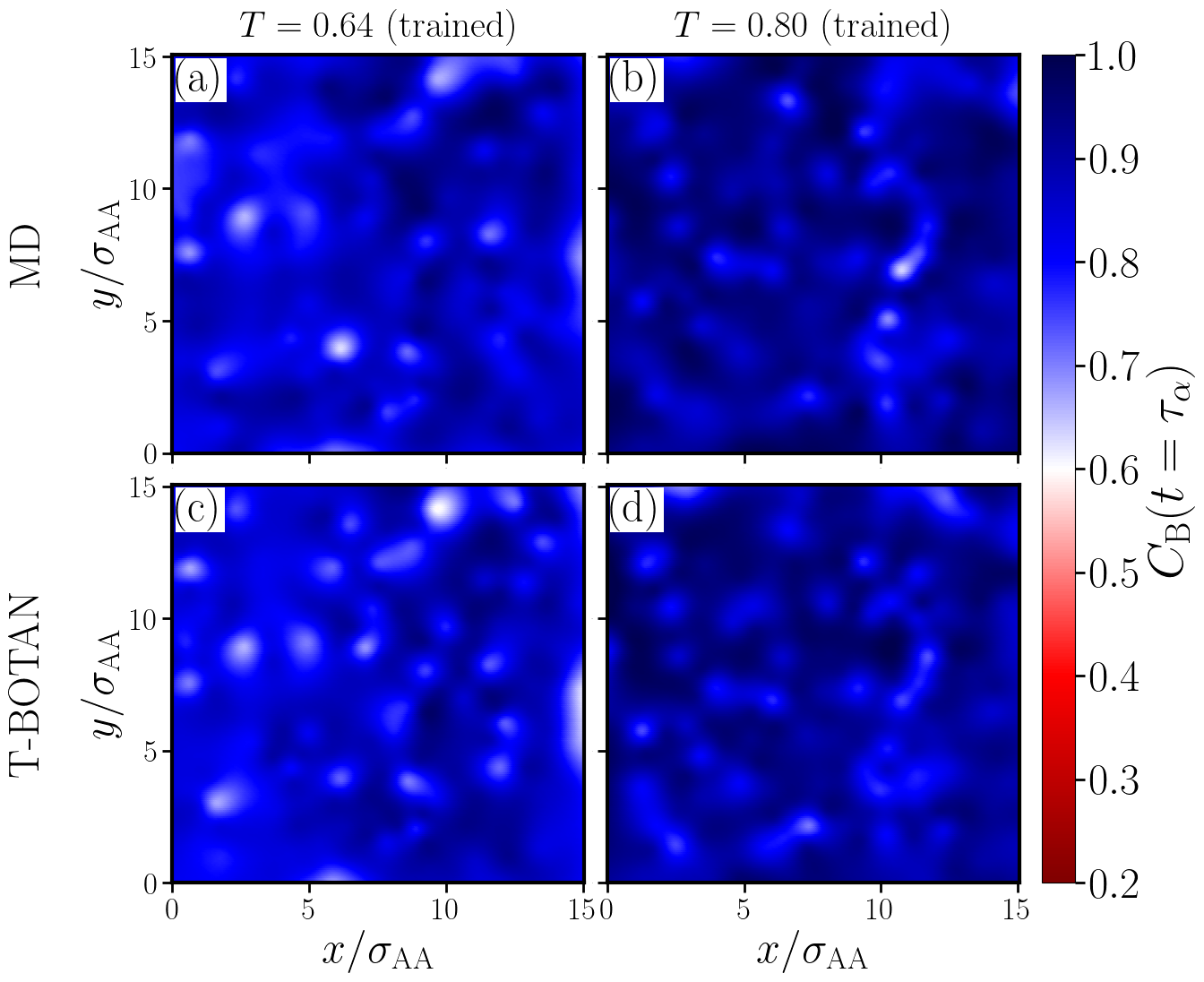}
    \caption{Comparison between the MD simulation data ((a) and (b)) and the T-BOTAN predictions ((c) and (d)) for the spatial distribution of $C_{\RM{B}}(t=\tau_{\alpha})$ at $T = 0.64$ and $0.80$.
    All data are shown for the same three-dimensional particle configuration, displayed as a cross section in the range $11.1 < z < 11.9$.}
    \label{fig:S_CB}
\end{figure}

\begin{figure}
    \centering
    \includegraphics[width=0.45\textwidth]{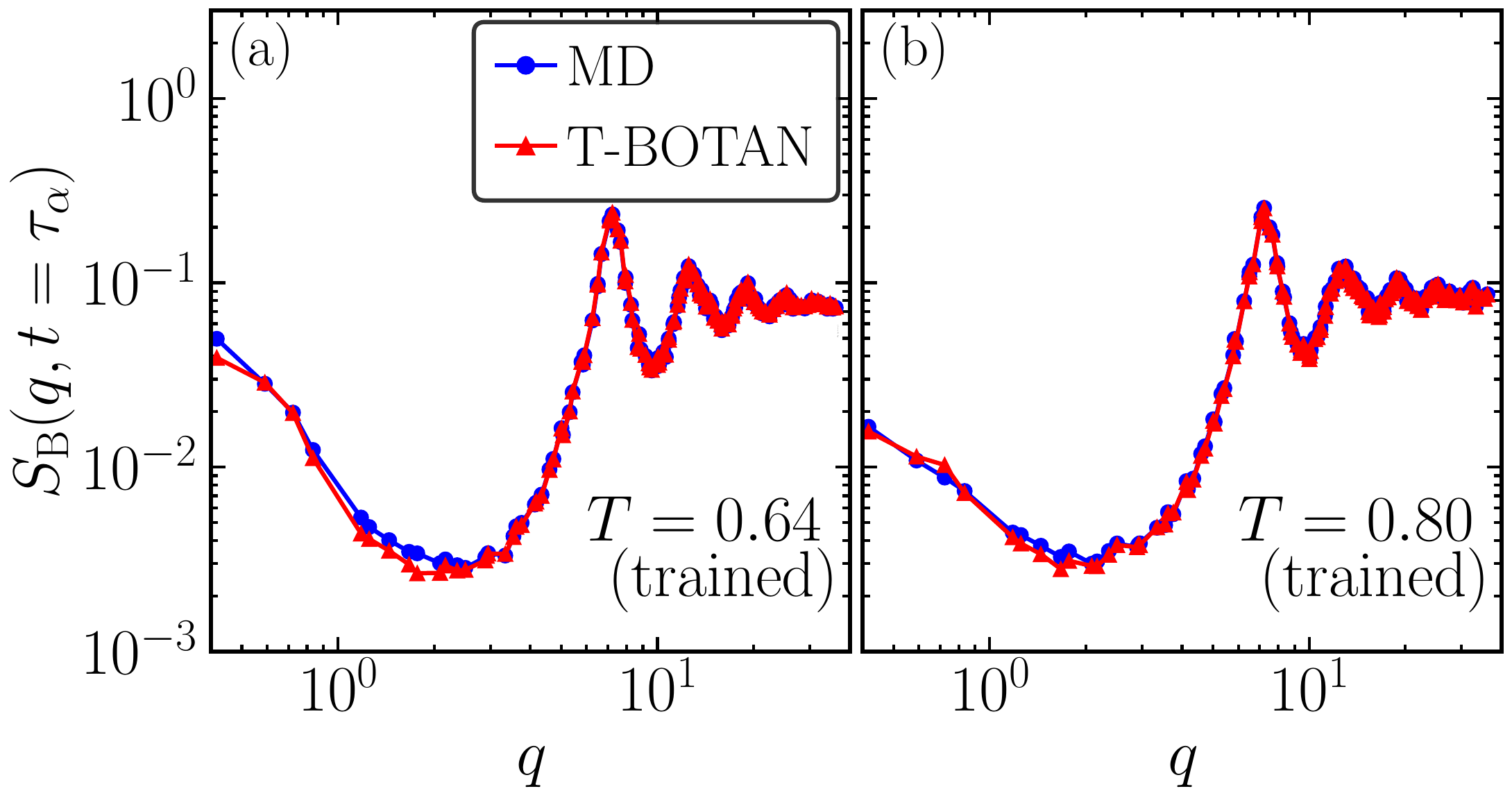}
    \caption{$S_{\RM{B}}(q,t=\tau_{\alpha})$ at $T = 0.64$ and $0.80$.}
    \label{fig:S_SB}
\end{figure}

\begin{figure}
    \centering
    \includegraphics[width=0.4\textwidth]{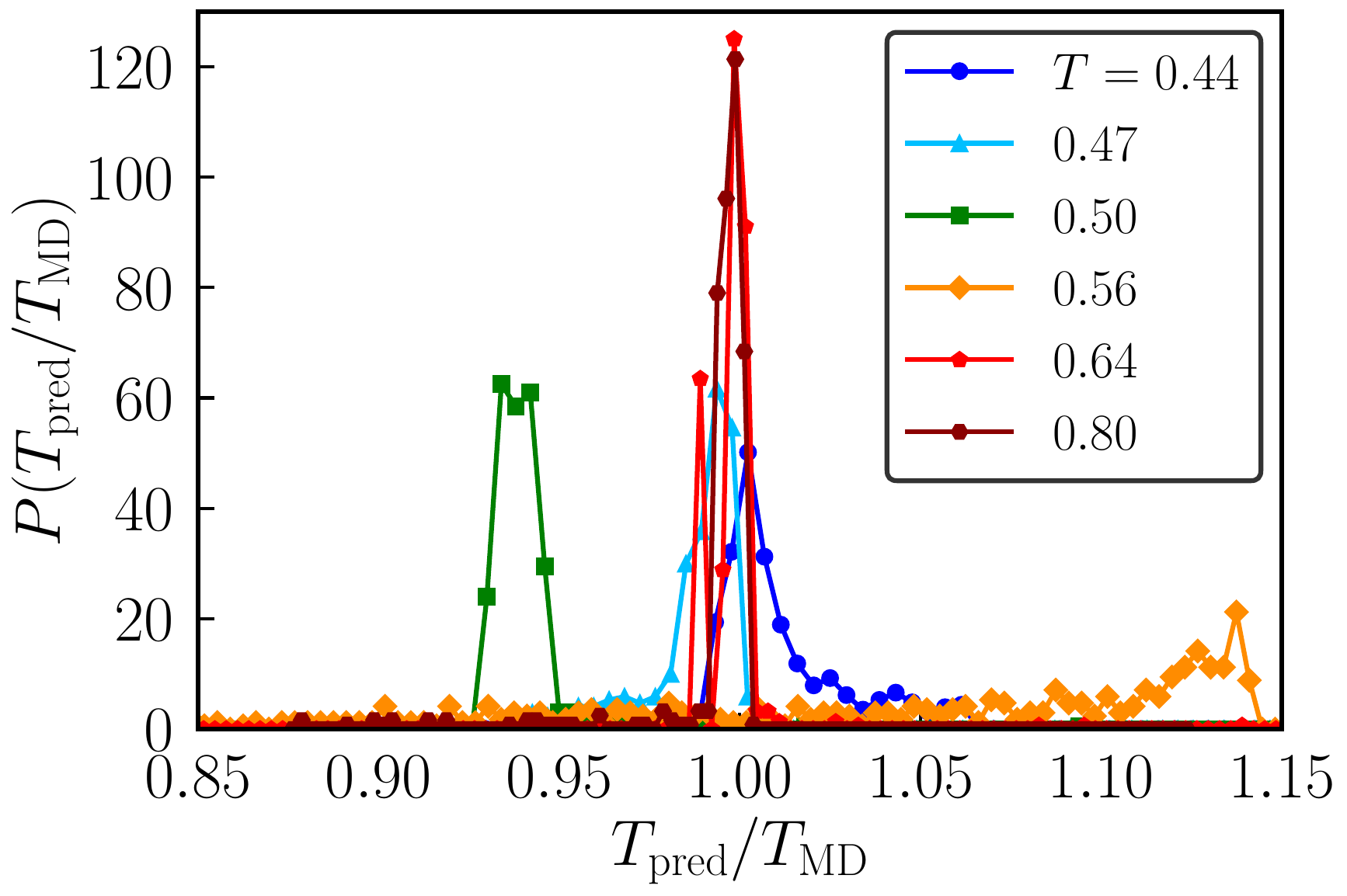}
    \caption{Distribution function $P(T_{\RM{pred}}/T_{\RM{MD}})$ for various $T$. 
    The distribution is normalized according to Eq.~(\ref{eq:P_of_T}).}
    \label{fig:S_Tpred_dist}
\end{figure}
Figure~\ref{fig:S_CB} shows typical snapshots of $C_{\RM{B}}(t=\tau_{\alpha})$ obtained from the MD simulations and the T-BOTAN predictions for $T = 0.64$ and $0.80$.
The correlation between them remains modest in the high-temperature regime, reflecting the increased randomness of particle motion, yet T-BOTAN successfully captures the qualitative spatial patterns. 
Figure~\ref{fig:S_SB} presents $S_{\RM{B}}(q,t=\tau_{\alpha})$ for $T = 0.64$ and $0.80$.
The $S_{\RM{B}}(q)$ obtained from the T-BOTAN predictions is almost identical to that from the MD simulations.
These results verify that T-BOTAN can reproduce dynamic heterogeneity in the high-temperature regime even without explicit temperature input.

Figure~\ref{fig:S_Tpred_dist} shows the distribution function $P(T_{\RM{pred}}/T_{\RM{MD}})$ for various $T$, where $T_{\RM{pred}}$ and $T_{\RM{MD}}$ denote the predicted temperature and the MD simulation temperature, respectively.
The distribution is normalized as
\begin{equation}
    \int \dd\left(\frac{T_{\RM{pred}}}{T_{\RM{MD}}}\right)
    P\left(\frac{T_{\RM{pred}}}{T_{\RM{MD}}}\right) = 1.
    \label{eq:P_of_T}
\end{equation}
For the trained temperatures ($T = 0.44, 0.47, 0.64$, and $0.80$), $P(T_{\RM{pred}}/T_{\RM{MD}})$ exhibits a sharp peak around $T_{\RM{pred}}/T_{\RM{MD}} = 1$, indicating that T-BOTAN accurately predicts the temperature of the input structures.
For the untrained temperatures ($T = 0.50$ and $0.56$), $P(T_{\RM{pred}}/T_{\RM{MD}})$ does not exhibit a sharp peak at $T_{\RM{pred}}/T_{\RM{MD}} = 1$.
When $T = 0.50$, the distribution attains its maximum around $T_{\RM{pred}}/T_{\RM{MD}} \approx 0.94$, implying that $T_{\RM{pred}} \approx 0.47$.
For $T = 0.56$, a small peak appears around $T_{\RM{pred}}/T_{\RM{MD}} \approx 1.14$, suggesting that T-BOTAN identifies the structure at $T = 0.56$ as being closer to that at $T = 0.64$.
This behavior indicates that T-BOTAN effectively projects structures at untrained temperatures onto those at the nearest trained temperatures.
These results are consistent with those presented in the main text.

\section{Hyperparameter \texorpdfstring{$\lambda_{\RM{g}}$}{pg} dependence}
\label{sec:S3}

\begin{figure}
    \centering
    \includegraphics[width=0.48\textwidth]{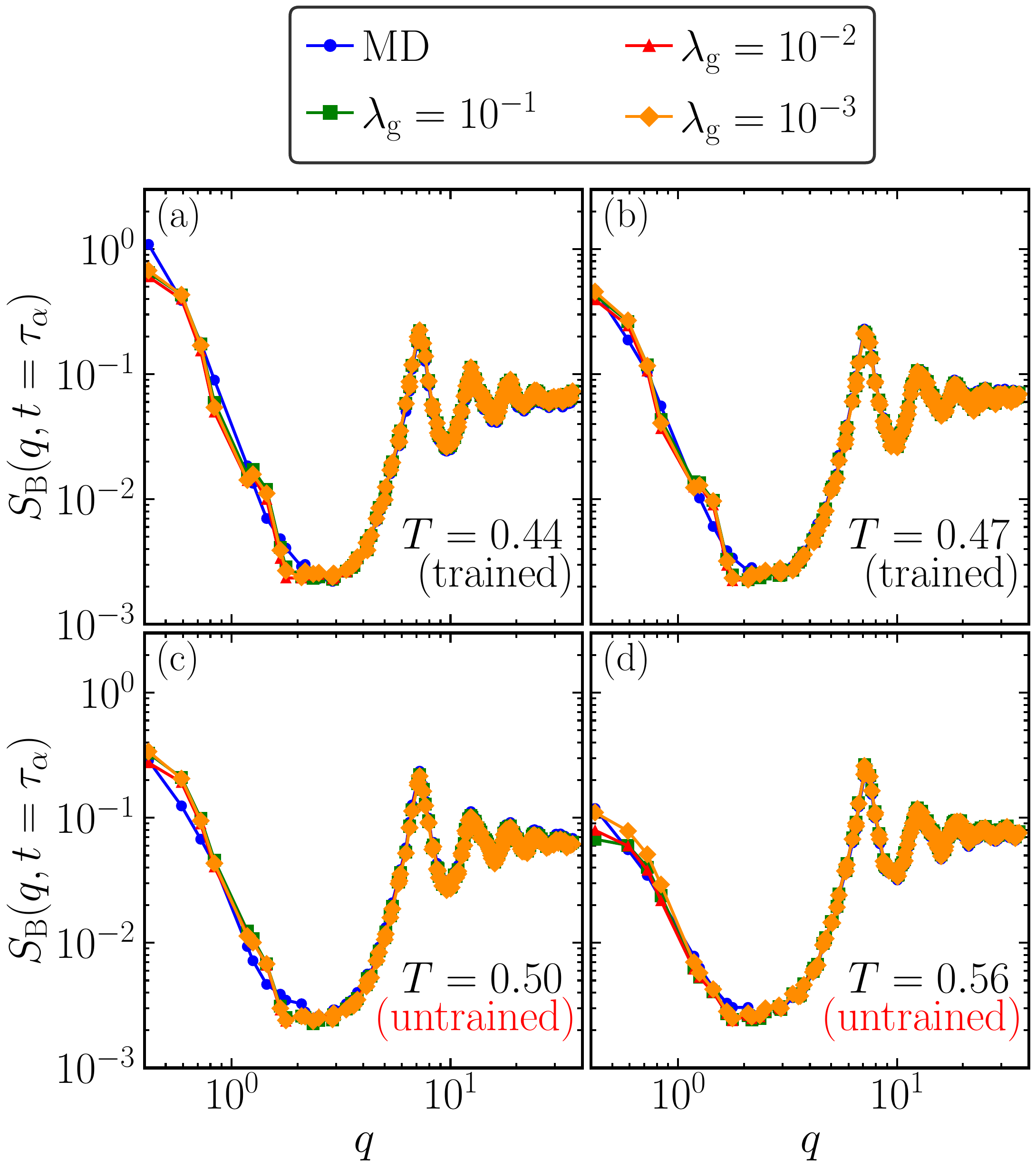}
    \caption{$\lambda_{\RM{g}}$ dependence of $S_{\RM{B}}(q,t=\tau_{\alpha})$ for various $T$.
    The data for $\lambda_{\RM{g}} = 10^{-2}$ are the same as those shown in the main text.}
    \label{fig:S_pg_SB}
\end{figure}

\begin{figure}
    \centering
    \includegraphics[width=0.35\textwidth]{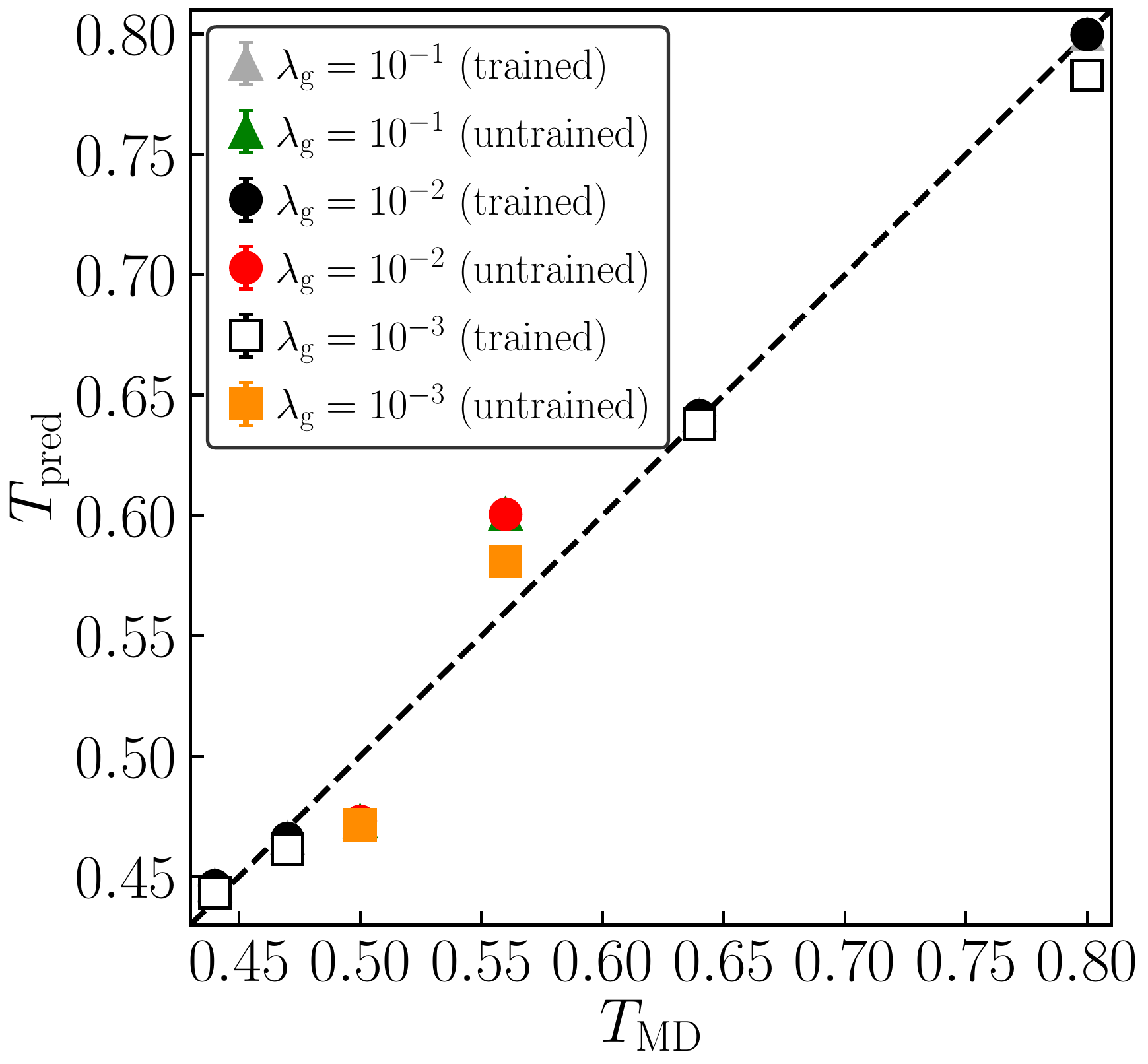}
    \caption{Comparison between $T_{\RM{pred}}$ and $T_{\RM{MD}}$ with standard errors for various $\lambda_{\RM{g}}$.
    Filled symbols represent temperatures included in the training dataset, whereas open symbols denote temperatures not used for training.
    The data for $\lambda_{\RM{g}} = 10^{-2}$ are the same as those shown in the main text.
    Error bars represent the standard deviation over different samples, but are smaller than the symbol size and thus not visible.}
    \label{fig:S_pg_Tpred}
\end{figure}
Next, we examine the dependence on the hyperparameter $\lambda_{\RM{g}}$, which controls the weight of the global loss term $\mathcal{L}_{\mathrm{global}}$.
Figure~\ref{fig:S_pg_SB} shows $S_{\RM{B}}(q,t=\tau_{\alpha})$ for various $T$ and $\lambda_{\RM{g}}$.
We find that $S_{\RM{B}}(q)$ is almost independent of $\lambda_{\RM{g}}$.

Figure~\ref{fig:S_pg_Tpred} presents a comparison between $T_{\RM{pred}}$ and $T_{\RM{MD}}$ for various $\lambda_{\RM{g}}$.
This figure also indicates that $T_{\RM{pred}}$ is essentially independent of $\lambda_{\RM{g}}$.

This weak dependence arises because $\mathcal{L}_{\mathrm{global}}$ is much smaller than the other loss terms, as shown in Fig.~\ref{fig:S_loss_all}, and therefore does not significantly influence the training process.

\section{Comparison of predictions based on inherent and thermal structures}
\label{sec:S4}
\begin{figure*}
    \centering
    \includegraphics[width=0.95\textwidth]{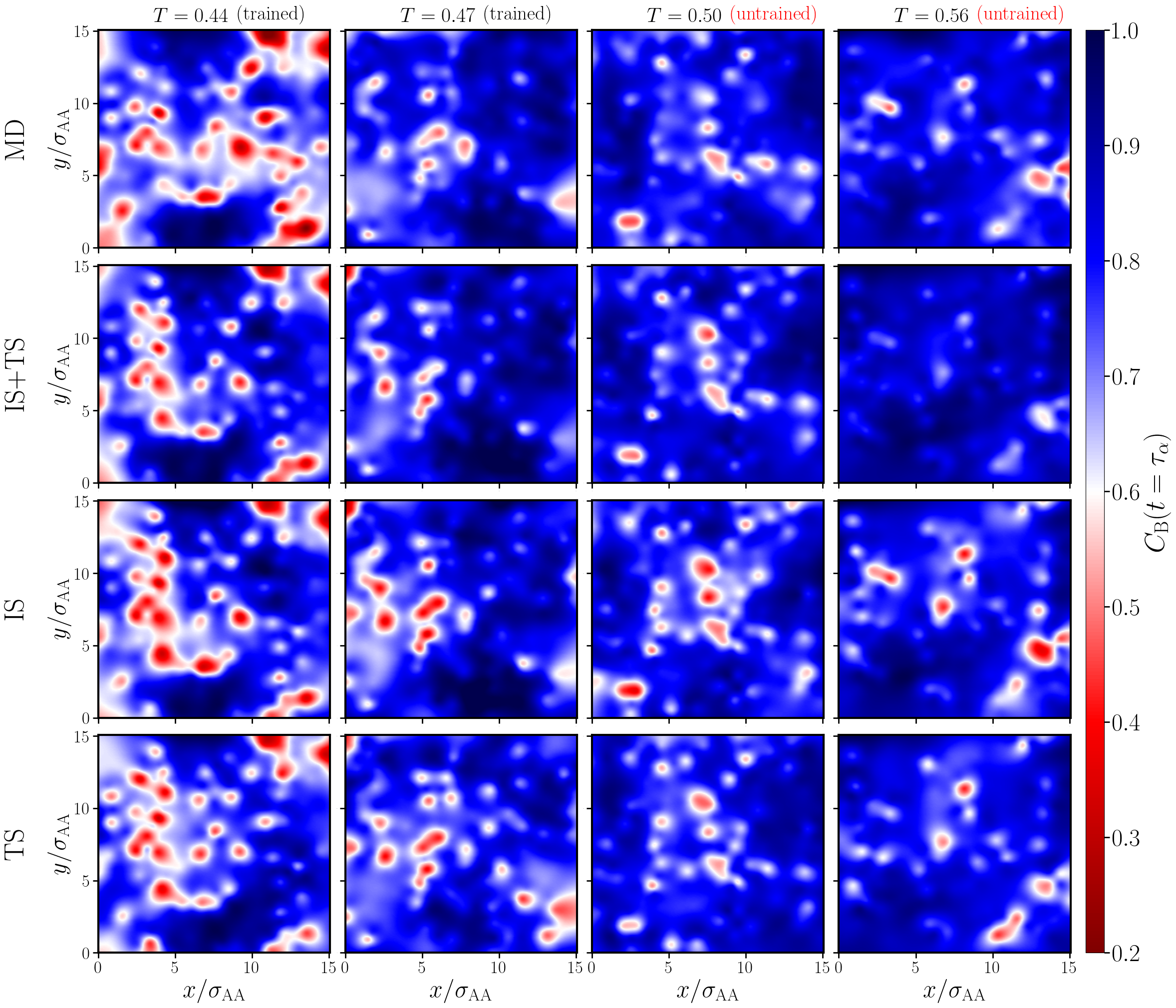}
    \caption{Comparison between MD simulation data and T-BOTAN predictions for the spatial distribution of $C_{\RM{B}}(t=\tau_{\alpha})$ for $T=0.44, 0.47, 0.50$, and $0.56$.
    ``IS+TS'' denotes the prediction obtained using both inherent and instantaneous thermal structures (as in the main text).
    ``IS'' and ``TS'' denote predictions based solely on the inherent structures and instantaneous thermal structures, respectively.
    All data are plotted for the same 3D particle configuration, in the cross section $11.1<z<11.9$.}
    \label{fig:S_CB_IS_vs_TS}
\end{figure*}
\begin{figure}
    \centering
    \includegraphics[width=0.48\textwidth]{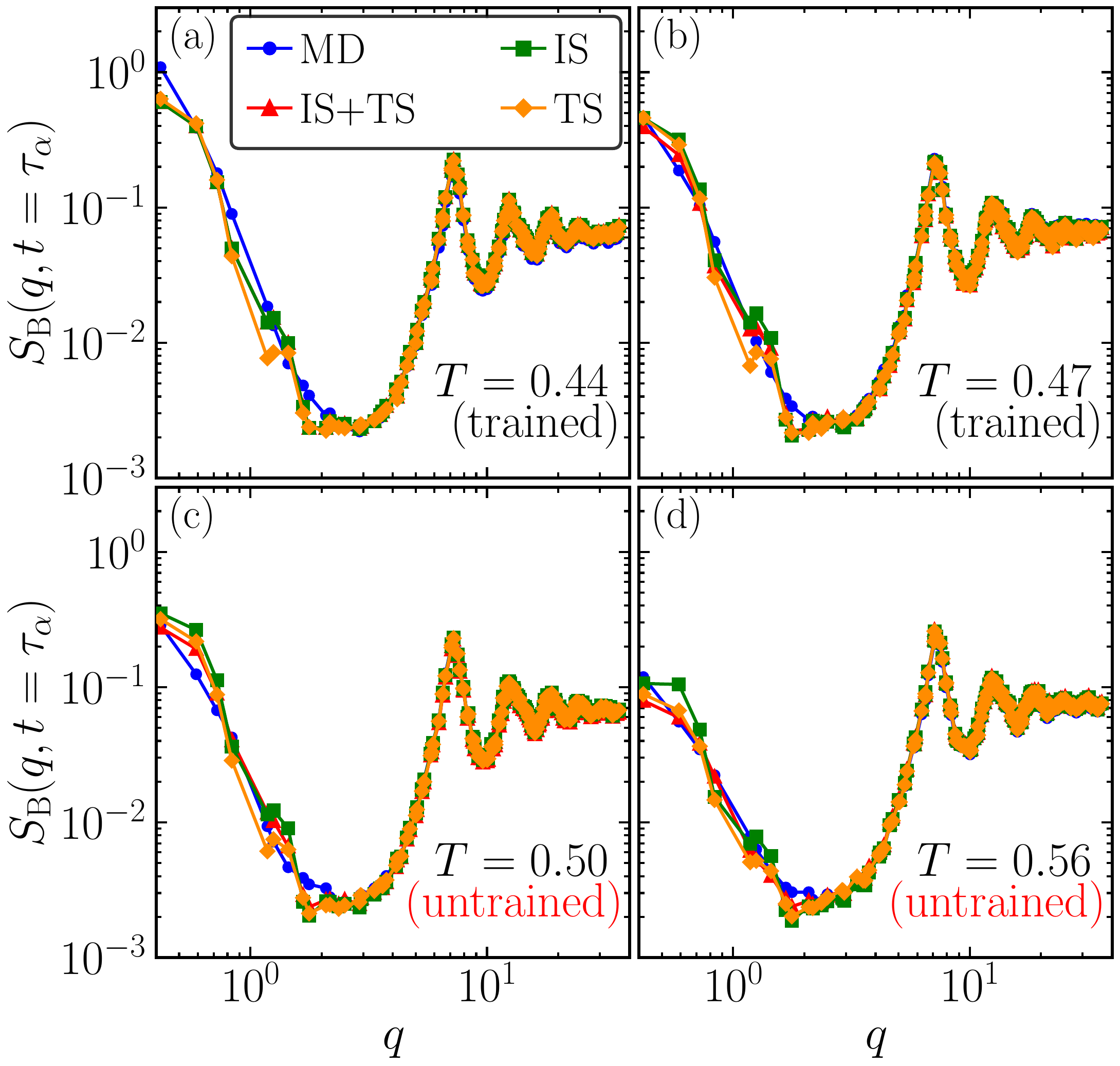}
    \caption{$S_{\RM{B}}(q,t=\tau_{\alpha})$ for various $T$. 
    ``IS+TS'' denotes the prediction obtained using both inherent and instantaneous thermal structures (as in the main text).
    ``IS'' and ``TS'' denote predictions based solely on the inherent structures and instantaneous thermal structures, respectively.
    (a) $T=0.44$ (b) $T=0.47$ (c) $T=0.50$ (d) $T=0.56$}
    \label{fig:S_SB_IS_vs_TS}
\end{figure}
\begin{figure}
    \centering
    \includegraphics[width=0.35\textwidth]{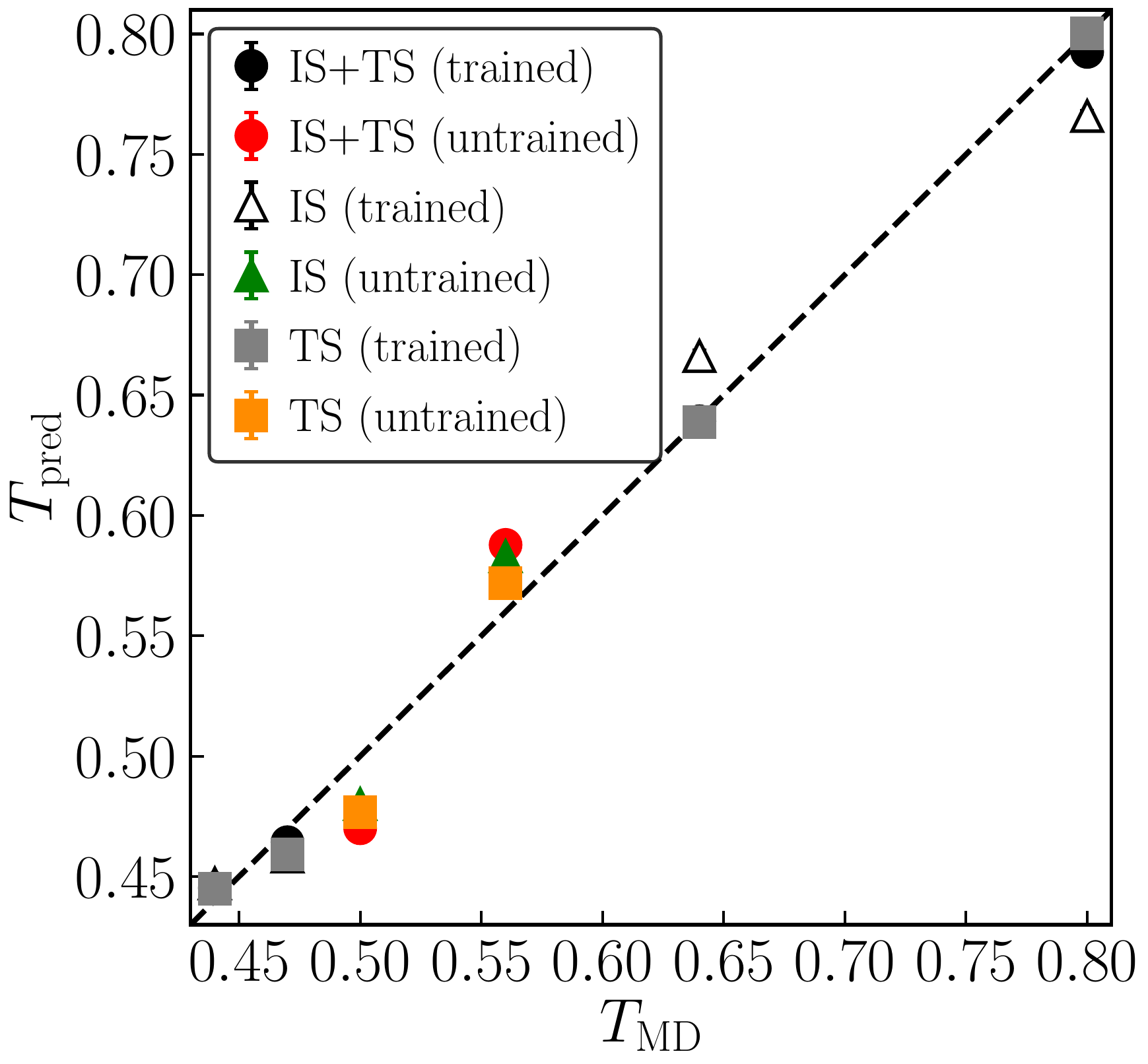}
    \caption{$T_{\RM{pred}}$ vs $T_{\RM{MD}}$ with standard error.
    ``IS+TS'' denotes the prediction obtained using both inherent and instantaneous thermal structures (as in the main text).
    ``IS'' and ``TS'' denote predictions based solely on the inherent structures and instantaneous thermal structures, respectively.
    Error bars represent the standard deviation over different samples, but are smaller than the symbol size and thus not visible.}
    \label{fig:S_Tpred_vs_TMD_IS_vs_TS}
\end{figure}
Finally, we present the results obtained using only ISs or only thermal structures as inputs.
Figure~\ref{fig:S_CB_IS_vs_TS} compares the particle configurations from the MD simulations and the corresponding T-BOTAN predictions, colored by $C_{\RM{B}}(t=\tau_{\alpha})$, for various temperatures.
Overall, the spatial distributions predicted from only inherent or only thermal structures remain qualitatively similar to those obtained using both representations.
However, the Pearson correlation coefficient between the predictions and the MD data is maximized when both inherent and thermal configurations are used as input (not shown), indicating that the combined structural information provides the most complete representation for dynamical prediction.

Figure~\ref{fig:S_SB_IS_vs_TS} presents $S_{\RM{B}}(q,t=\tau_{\alpha})$ for various temperatures.
Consistent with the spatial maps, the predictive performance for dynamic heterogeneity using only inherent or only thermal structures is largely comparable to that obtained using both inputs, suggesting that each structural representation individually retains substantial information relevant to the collective dynamics.

Figure~\ref{fig:S_Tpred_vs_TMD_IS_vs_TS} shows the comparison between $T_{\RM{pred}}$ and $T_{\RM{MD}}$.
When only inherent structures are provided, the predicted temperature corresponds to the equilibrium temperature at which the configurations were prepared (i.e., the parent temperature prior to quenching).
When only thermal structures are used, the prediction reflects the thermodynamic temperature associated with the equilibrated configurations.
In both cases, the temperature is obtained as a data-driven statistical estimate from structural descriptors.
In the high-temperature region, the accuracy based solely on inherent structures slightly decreases, even for temperatures included in the training set.
These trends indicate that thermal structural features contribute more strongly to temperature discrimination at elevated temperatures, whereas inherent structures alone become less distinct as temperature increases.
Overall, the results confirm that temperature-dependent structural correlations are present in both representations and can be captured by the learned model.

\bibliography{ref}


\end{document}